\documentclass[3p,twocolumn]{elsarticle}

\usepackage{hyperref}
\usepackage{textcomp,gensymb,amsmath,amsthm,amssymb,amsfonts}
\usepackage{siunitx}
\usepackage{multirow,subscript}
\usepackage{color}

\newcommand{\sampleMn}{Mn\textsubscript{0.25}TaS\textsubscript{2}}
\newcommand{\sampleFe}{Fe\textsubscript{0.25}TaS\textsubscript{2}}
\newcommand{\host}{TaS\textsubscript{2}}

\usepackage{graphicx}

\makeatletter
\def\ps@pprintTitle{%
     \let\@oddhead\@empty
     \let\@evenhead\@empty
     \def\@oddfoot{\footnotesize\itshape\hfill\today}%
     \let\@evenfoot\@oddfoot}
\makeatother
\renewcommand*{\today}{December 10, 2015}

\biboptions{sort&compress}

\begin{document}

\begin{frontmatter}

\title{Structural and Magnetic Characterization of Large Area, Free-Standing Thin Films of Magnetic Ion Intercalated Dichalcogenides \sampleMn{} and \sampleFe{}}

\author[goe]{Th.~Danz}
\author[rug]{Q.~Liu}
\author[HMFL]{X.\,D.~Zhu}
\author[Pohang]{L.\,H.~Wang}
\author[Pohang,Rut]{S.\,W.~Cheong}
\author[tub,hzb]{I.~Radu}
\author[goe]{C.~Ropers}
\author[rug]{R.\,I.~Tobey\corref{cor}}
\ead{r.i.tobey@rug.nl}

\cortext[cor]{Corresponding author}
\address[goe]{4th Physical Institute, Solids and Nanostructures, University of G\"ottingen, 37077 G\"ottingen, Germany}
\address[rug]{Zernike Institute for Advanced Materials, University of Groningen, 9747 AG Groningen, The Netherlands}
\address[HMFL]{High Magnetic Field Laboratory, Chinese Academy of Sciences, Hefei 230031, P.\,R.\,China}
\address[Pohang]{Laboratory for Pohang Emergent Materials and Max Planck POSTECH Center for Complex Phase Materials, Pohang University of Science and Technology, Pohang 790-784, Korea}
\address[Rut]{Rutgers Center for Emergent Materials and Department of Physics and Astronomy, Rutgers University, New Brunswick, NJ 08901, USA}
\address[tub]{Institute for Optics and Atomic Physics, Technical University of Berlin, 10623 Berlin, Germany}
\address[hzb]{Helmholtz-Zentrum Berlin, BESSY II, 12489 Berlin, Germany}

\begin{abstract}
Free-standing thin films of magnetic ion intercalated transition metal dichalcogenides are produced using ultramicrotoming techniques. Films of thicknesses ranging from \SI{30}{nm} to \SI{250}{nm} were achieved and characterized using transmission electron diffraction and X-ray magnetic circular dichroism. Diffraction measurements visualize the long range crystallographic ordering of the intercalated ions, while the dichroism measurements directly assess the orbital contributions to the total magnetic moment. We thus verify the unquenched orbital moment in \sampleFe{} and measure the fully quenched orbital contribution in \sampleMn. Such films can be used in a wide variety of ultrafast X-ray and electron techniques that benefit from transmission geometries, and allow measurements of ultrafast structural, electronic, and magnetization dynamics in space and time.
\end{abstract}


\end{frontmatter}

\section{Introduction}
Transition metal dichalcogenides (TMDCs) are a broad class of layered materials with a variety of structural and electronic properties. Depending on the specific compound and layer stacking, TMDCs can exhibit metallic (e.g. TaS\textsubscript{2}), semi-metallic (e.g. WTe\textsubscript{2}), semiconducting (e.g. MoSe\textsubscript{2}), or insulating (e.g. HfS\textsubscript{2}) behaviour \cite{Wilson1969, Wang2012}. Recent work has focused on the emergence and dynamics of charge density wave formation (commensurate and incommensurate) concomitant with metal-to-insulator transitions \cite{Dean2011,Hellmann2012}. Yet other work dealt with the properties of monolayers of these materials. Loose interlayer bonding via van der Waals forces allows the fabrication of single monolayers of material which often exhibit electronic phases that are starkly different from bulk properties. Studies on monolayers have brought forth possibilities in optoelectronics and valleytronics \cite{Huang2013,Jariwala2014}, and spintronics \cite{Zhu2011}.

The intrinsic properties of the bulk materials can be further expanded by \textit{intercalation} of atoms and small molecules between the layers. In particular, the incorporation of 3\textit{d} transition metals results in the onset of an array of magnetic properties which can be tuned by intercalation concentration, choice of intercalated species, and choice of host lattice \cite{Parkin1980}. In total, the range of magnetic properties is enormous, providing a platform on which to study magnetism and domain structure as well as their dynamics.   

Here we discuss a sample preparation technique that provides free-standing thin films of magnetic ion intercalated \host{}. In thin film form, these samples open the possibility for a number of magnetization dynamics studies with new techniques such as ultrafast electron diffraction \cite{Eichberger2010, Zhu2013} and microscopy \cite{Sun2015}, as well as X-ray absorption \cite{Mohr-Vorobeva2011}, all of which benefit from transmission geometries where the penetration depth of both pump and probe can be easily matched. Moreover, for X-ray absorption studies, the transmission geometry is the most robust when compared to its total electron yield (TIY) or fluorescence yield (FY) counterparts in that it provides a direct measure of dichroic absorption and is unaffected by electronic saturation effects \cite{Nakajima}. The samples were prepared using ultramicrotoming techniques \cite{Malis1990}, and their structural and magnetic properties were measured using transmission electron diffraction and X-ray circular magnetic dichroism, respectively. Two different materials were studied, \sampleMn{} and \sampleFe{}, which differ in their magnetic properties while sharing the same structure.

\section{Material Description}
TMDCs (chemical formula MX$_2$ where X is a chalcogen and M a transition metal) consist of stacks of X--M--X sandwich layers which are van der Waals bonded along the $c$-axis \cite{Revelli1979}. Variations in layer stacking result in a wide range of physical properties, while particular stacking configurations promote intercalation. For \host{}, the metallic 2\textit{H} polytype facilitates intercalation.
\begin{figure}
    \includegraphics[width=1.0\columnwidth]{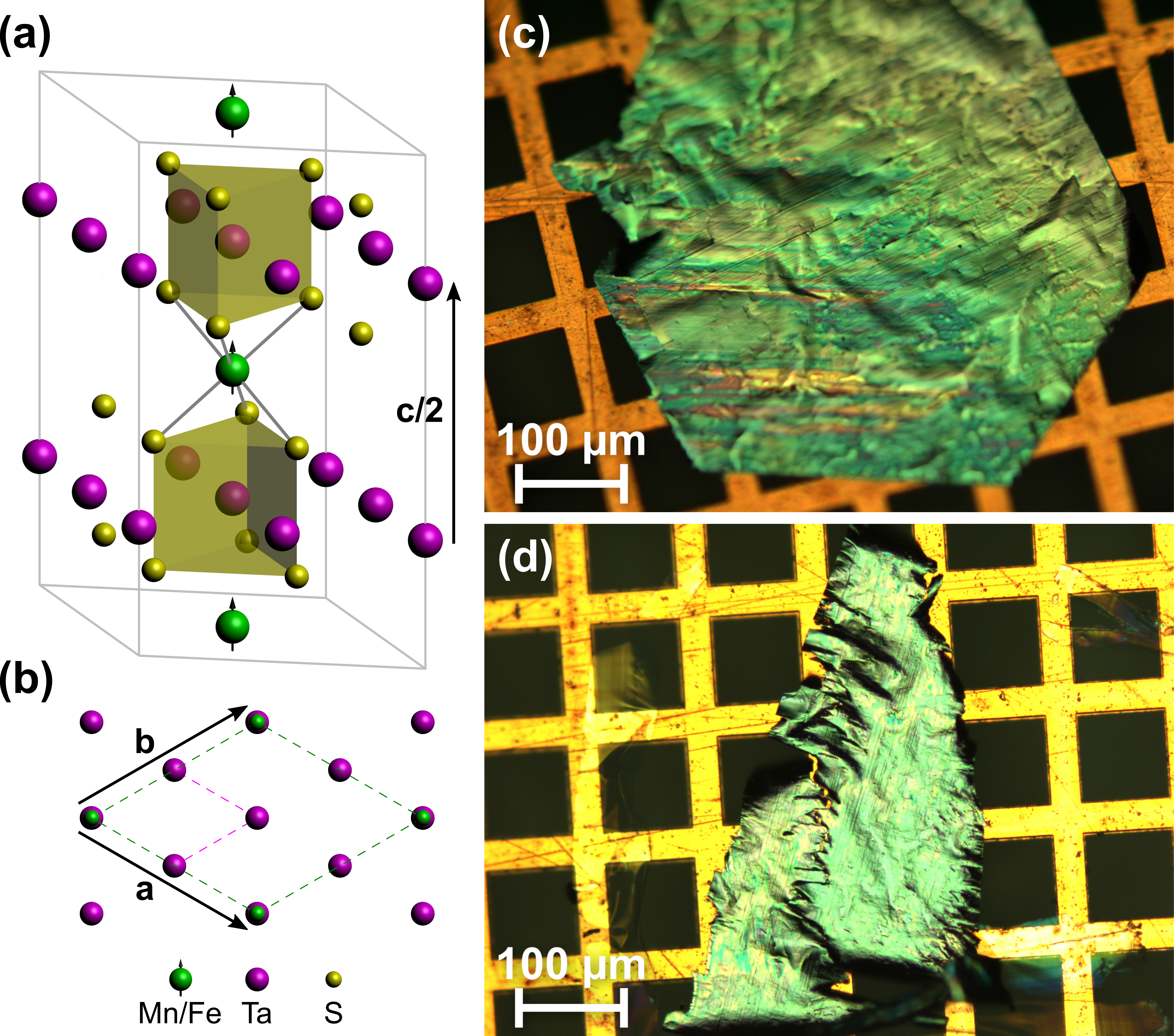}
    \caption{(a) Unit cell of 3\textit{d} ion intercalated TMDC samples. Trilayers of tantalum and sulfur are stacked along the $c$-axis. Magnetic ions fit between the layers in specific locations determined by the concentration $y$. (b) Planar view shows the crystallization of intercalants at $y = 0.25$. (c) \SI{100}{nm} sample of \sampleMn{}, and (d) \SI{200}{nm} sample of \sampleFe{} on a 200 lines per inch copper grid.}
    \label{fig:optical}
\end{figure}
Upon intercalation, the new chemical formula is A$_y$MX$_2$, which in the context of TMDCs is understood to maintain the MX$_2$ structure, while new ions are incorporated between the host layers. The process of 3\textit{d} ion intercalation occurs during the growth phase, for example using iodine vapor transport. Interestingly, particular intercalant concentrations lead to long-range crystalline order of intercalants between the host layers. This is the case for two situations, that of intercalant density $y = 1/4$ and $y = 1/3$, where the intercalants form an ordered $2\times2$ or $\sqrt{3}\times\sqrt{3}$ superlattice, respectively \cite{Vandenberg-Voorhoeve1976,Parkin1980,Hinode1995}. When such ordered states occur, the intercalated ion is known to reside in specific sites of the host lattice (the trigonal antiprismatic hollow site between the layers of 2\textit{H}-\host{}), ensuring that the local environment of the magnetic ion is unchanged as one varies the intercalant species or host lattice. A three-dimensional view of the crystallographic structure is shown in Fig.~\ref{fig:optical}(a) with specific emphasis on the location of the intercalant ion between X--M--X sheets. The crystallization of intercalants is best visualized in a planar view along the $c$-axis, as shown in Fig.~\ref{fig:optical}(b).

Bulk materials are characterized by Curie temperatures of \SI{80}{K} for \sampleMn{} \cite{Parkin1980} and \SI{155}{K} \cite{Vannette2009} for \sampleFe{}. The large intercalant-intercalant distance inhibits direct exchange mechanisms, and an indirect RKKY interaction is understood to play a crucial role \cite{Parkin1980,Ko2011}. In this case, the exchange interaction between local moments is mediated by the conduction electrons of the host lattice.  The isolated ions thus exhibit structural aspects reminiscent of local-moment magnetic systems.  At the same time, there are indications of itinerant magnetism in \sampleMn{} due to a strong hybridisation between the intercalant's 3\textit{d} states with the Ta 5\textit{d\textsubscript{z\textsuperscript{2}}} states which form the conduction band \cite{Motizuki1992}. This class of materials thus bridges the divide between truly local-moment magnetism (as in 4\textit{f} systems) and itinerant ferromagnetism observed in 3\textit{d} transition metals.

The two materials are further distinguished by their magnetic orientation.  While \sampleMn{} is an easy plane ferromagnet with low coercivity \cite{Onuki1986}, \sampleFe{} exhibits a strong out-of-plane magnetic anisotropy with high coercivity \cite{Morosan2007}. Additionally, as we will confirm, \sampleFe{} is known to have a large orbital contribution to the magnetization, while the orbital contribution in \sampleMn{} is fully quenched.

\section{Preparation of Free-Standing Thin Films}
TMDCs can be readily thinned down to monolayer thickness using the `scotch tape method' \cite{Wilson1969}, as well as by lithium intercalation \cite{Zeng2011} or by ultrasonification \cite{Coleman2011}.  However, due to charge transfer from the intercalant to the host and a concomitant change of the interlayer bonding towards a covalent bond \cite{Liang1986}, standard cleaving methods must be abandoned in the case of 3\textit{d} ion intercalated TMDCs. To our knowledge, there have been no reports of exfoliation of thin samples of \sampleMn{} and \sampleFe{}.

Recently, the preparation of free-standing films of 1\textit{T}-\host{} using an ultramicrotome was reported \cite{Eichberger2013}. The resulting films had lateral dimensions of \SI{\sim 200}{\micro m} and thickness of \SI{30}{nm}. On the basis of supporting transmission electron microscope (TEM) diffraction images, it has been shown that the crystallinity of the sample and the transition temperatures between the different charge density wave phases are unaffected, and that the preparation principle is suitable for making thin TMDC samples. Following this demonstration, we applied similar methodologies, however here to the more strongly bonded 3\textit{d} ion intercalated TMDCs.

Following standard techniques used in specimen sectioning \cite{Malis1990}, we embedded \sampleMn{} and \sampleFe{} bulk crystals with lateral dimensions of \SI{\sim 500}{\micro m} in transparent epoxy (Epofix Cold-Setting Embedding Resin) for cutting in a Leica EM UC6 ultramicrotome. After mixing resin and hardener, the epoxy was heated for a few minutes in an oven at \SI{60}{\degreeCelsius} in order to lower its viscosity and remove air bubbles. Afterwards, we covered the bottom of $\SI{15}{mm}\times\SI{7}{mm}$-sized molds with a thin epoxy film and cured the epoxy for \SI{\sim 30}{min} at \SI{60}{\degreeCelsius}. Then we cut scratches in the epoxy layers using a razor blade, introduced bulk crystals in an upright position and filled the rest of the molds with epoxy. After another \SI{\sim 3}{h} of curing at \SI{60}{\degreeCelsius}, the epoxy blocks were ready for sectioning.  Epoxy blocks were trimmed to expose the \textit{ab}-plane of the sample material. After pre-sectioning with a glass knife, a diamond knife (dEYEmond ULTRA \SI{45}{\degree}) was used to produce thin sections from the block face. These were transferred onto square mesh copper TEM grids using a metallic loop suspending the sections in a thin water film (`perfect loop').  As read out from the ultramicrotome settings, we were able to obtain large area sections from \SI{250}{nm} down to \SI{30}{nm} in thickness. We found that \sampleMn{} was more easily sectioned than \sampleFe{}, which had a tendency to disintegrate during cutting. We attribute this to deformations of the \sampleFe{} lattice induced during the clipping of a small fraction from a spatially more extended crystal platelet. For \sampleMn{} in contrast, we had suitably sized bulk crystals available from the start.

\section{Characterization}
\subsection{Optical Microscopy}
Optical microscopy was conducted for every sample, representative images of which are shown in Fig.~\ref{fig:optical}(c)~and~(d). The sectioned samples showed large lateral areas of several hundreds of microns for \sampleMn{} and slightly smaller for \sampleFe{}, and a good adhesion to the grids.

In the optical images, we can identify regions of the samples that appear grooved. This is most easily seen as horizontal lines near the bottom of Fig.~\ref{fig:optical}(c). The grooves correspond to the cutting direction of the ultramicrotome and indicate imperfections that develop during multiple uses of the diamond knife. In Ref.~\cite{Eichberger2013}, it is suggested that this is caused by a reaction of the tantalum with the diamond blade. In our studies, we did not use a virgin knife, and thus we are unable to classify the origin of the knife damage. However, over the course of many cuts, we note that the density of scratches and nicks in the blade did not noticeably increase, while the sample quality remained unchanged.

\subsection{Transmission Electron Diffraction}
\begin{figure}[t]
    \centering
    \includegraphics[width=\columnwidth]{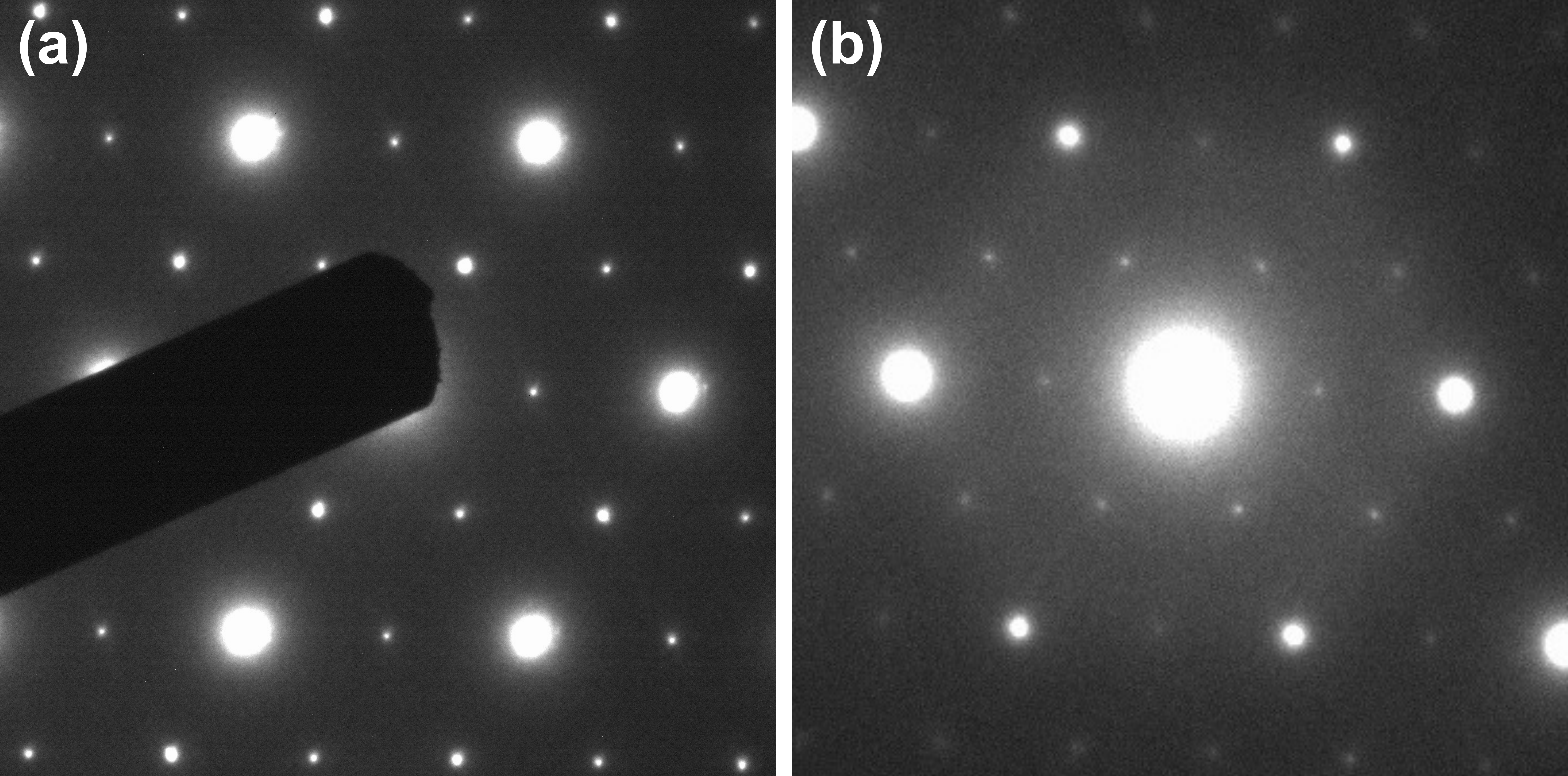}
    \caption{Selected area electron diffraction (SAED) images of (a) \SI{50}{nm} sample of \sampleMn{}, and (b) \SI{200}{nm} sample of \sampleFe. Weak superlattice spots are visible in both diffraction patterns halfway between the structural reflections of 2\textit{H}-\host. The images are scaled to accommodate different camera lengths.}
    \label{fig:tem}
\end{figure}

In order to verify the crystallinity of the sectioned samples, and particularly to investigate the ordering of intercalated ions, we recorded selected area electron diffraction (SAED) patterns in a TEM. Images recorded at \SI{200}{keV} electron energy are shown in Fig.~\ref{fig:tem}. For \sampleFe{}, a thinner edge region was required in order to clearly witness a nice diffraction pattern. Importantly, both diffraction patterns exhibit the expected superlattice reflections midway between the structural reflections of the host lattice, indicative of the $2\times2$ ordering of the intercalants. Furthermore, we observe no sign of a $\sqrt{3}\times\sqrt{3}$ superlattice which would emerge in case of intercalant excess or disorder \cite{Choi2009}. In general, good quality diffraction patterns were acquired on the thinnest samples (XMCD measurements discussed below can be performed on all sample thicknesses), while \sampleMn{} showed better spatial homogeneity in intercalant concentration than \sampleFe{}.

\subsection{XMCD Spectroscopy}
To assess the magnetic properties of the thinned samples, X-ray magnetic circular dichroism (XMCD) measurements in a transmission geometry were performed using the ALICE chamber at beamline PM3 of the BESSY II synchrotron at Helmholtz-Zentrum Berlin. The ALICE chamber features a liquid helium cooled sample holder and an electromagnet, both of which rotate with respect to the incoming beam direction and facilitate the study of both out-of-plane and in-plane magnetization characteristics. Details on the broader versatility of ALICE can be found in Ref.~\cite{Abrudan2015}. Measurements were conducted at a range of temperatures below the respective Curie temperatures of the two samples. Full spectroscopic information over the $L_{3,2}$ absorption edges of Fe and Mn was collected. Since XMCD is sensitive to the magnetization component parallel to the X-ray wave vector, measurements of \sampleFe{} were performed at normal incidence, while \sampleMn{} (and the applied field) was rotated by an angle $\Theta$ with respect to the X-ray beam direction to maximize the projection of the magnetization onto the X-ray wave vector.  Rotation angles achieved did not exceed \SI{40}{\degree} from normal. For a fixed X-ray helicity, the direction of the magnetic field was switched, and the transmitted intensity was acquired for oppositely magnetized films, denoted as $I_+$ and $I_-$, from which we calculate the absorption cross sections $\mu_\pm=-\log\left(I_\pm/I_0\right)$ for the two magnetization directions ($I_0$ is the incident photon flux).
\begin{figure}
    \centering
	\includegraphics[width=1.0\columnwidth]{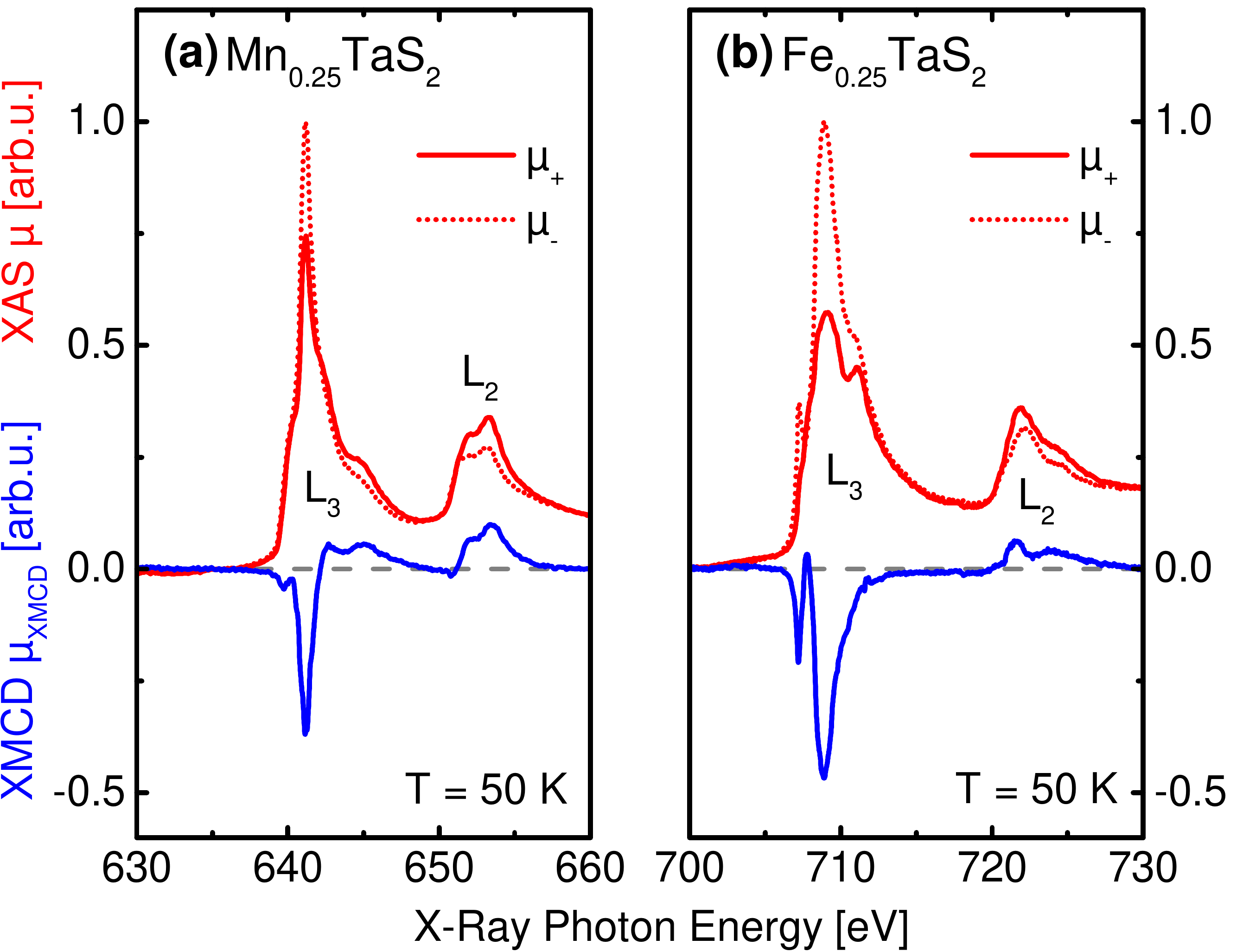}
    \caption{X-ray absorption spectra $\mu_\pm$ (red) for (a) \SI{150}{nm} sample of \sampleMn{}, and (b) \SI{200}{nm} sample of \sampleFe{}, for two orientations of applied magnetic field. The sample and applied field were rotated \SI{40}{\degree} with respect to the incoming beam for \sampleMn, while \sampleFe{} was acquired at normal incidence. XMCD spectra are shown in blue. The asymmetry in $L_3$ and $L_2$ for the case of \sampleFe{} indicates a large unquenched orbital moment.}
    \label{fig:xmcdxas}
\end{figure}

Figure~\ref{fig:xmcdxas} shows representative X-ray absorption spectra $\mu_\pm$ (red) for \sampleMn{} and \sampleFe{} across the respective $L_{3,2}$ edges for both orientations of the magnetic field. Included in the figure is the extracted XMCD signal $\mu_\text{\tiny XMCD}$ (blue) for the two materials, calculated following standard treatments for XMCD analysis \cite{OBrien1994, Chen1995}:
\begin{align}
\mu_\text{\tiny XMCD}(E)=\frac{1}{\gamma\cos\left(\Theta\right)}\left[\mu_+(E)-\mu_-(E)\right]
\end{align}
The XMCD signal has been corrected for both angle of incidence $\Theta$ and degree of helicity of the source $\gamma$. We used $\gamma \approx .925$ which is characteristic for the off-axis emission of the bend magnet radiation at beamline PM3 (specifically at the Fe $L$ edge) \cite{Kachel}.

A cursory comparison between the XMCD spectra for the two materials unveils a notable difference, namely, the integrated intensity of the $L_{3,2}$ edges are vastly different. As is well known from XMCD analysis, differences in the integrated intensity at the two edges is a manifestation of varying degrees of spin and orbital contributions to the total magnetization.

Applying sum rules to the respective edges, we can extract the ratio of orbital to spin magnetic moment \cite{Chen1995}:
\begin{align}
\frac{m_\text{orb}}{m_\text{spin}}=\frac{2q}{9p-6q}
\end{align}
where $p$ represents the area under the $L_3$ edge, and $q$ the overall area under both absorption edges:
\begin{align}
p&=\int_{L_3}\mu_\text{\tiny XMCD}(E)\,\text{d}E\\
q&=\int_{L_3+L_2}\mu_\text{\tiny XMCD}(E)\,\text{d}E
\end{align}
The measured ratio of orbital to spin magnetic moment for \sampleFe{} is found to be \SI{.39(4)}{} and in close agreement with previously obtained values of .33 \cite{Ko2011}.  Meanwhile the measured value for \sampleMn{} is \SI{-.06(6)}{}, which to our knowledge is the first direct measurement of the fully quenched orbital moment in this material. The results and corresponding error bars are derived from the ensemble of measurements below \SI{50}{K} and \SI{110}{K} for \sampleMn{} and \sampleFe{}, respectively.

\begin{figure}
    \centering
    \includegraphics[width=\columnwidth]{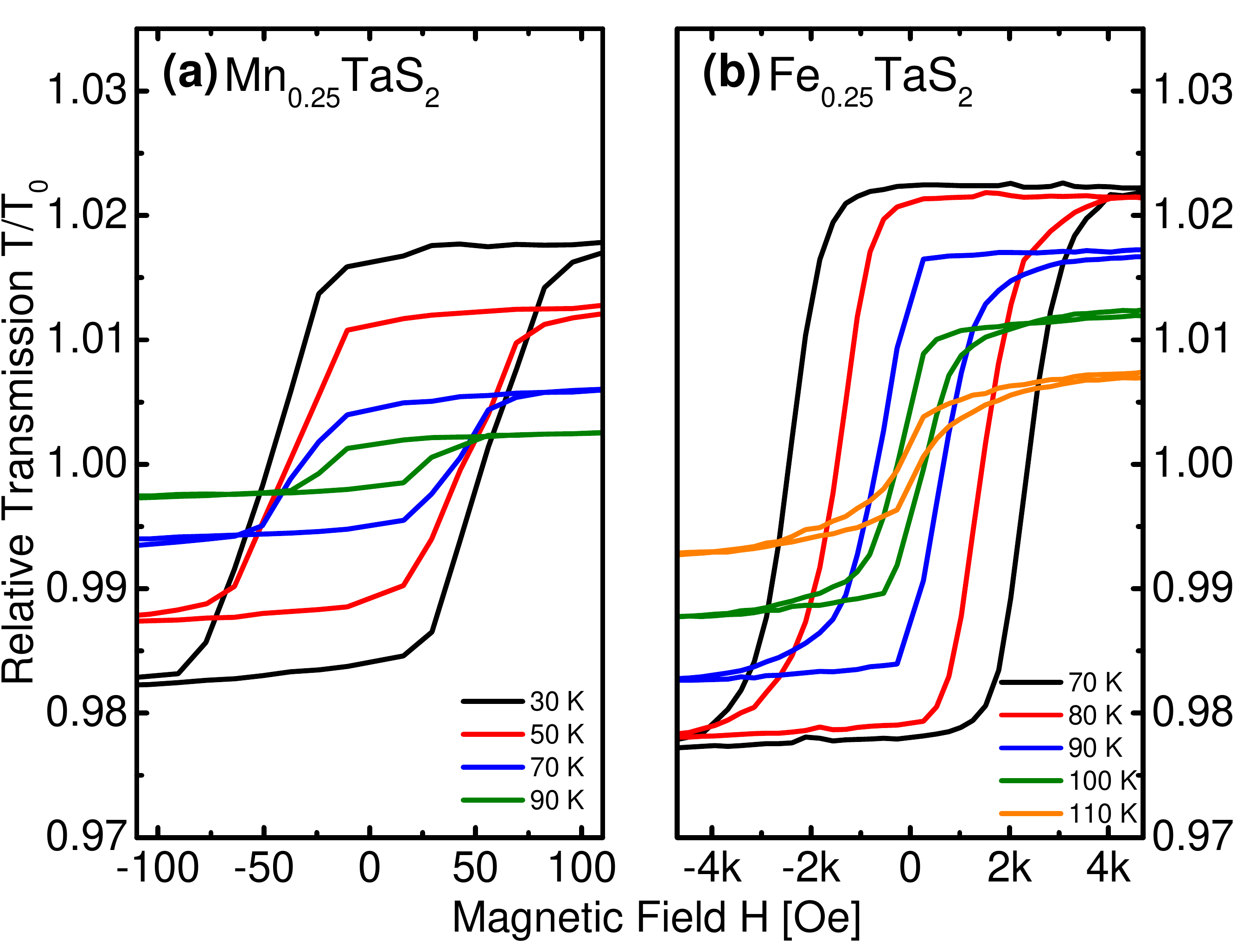}
    \caption{Hysteresis curves at the $L_3$ edge of (a) \SI{50}{nm} sample of \sampleMn{}, and (b) \SI{200}{nm} sample of \sampleFe{}. Note the different field strengths for the two samples.}
    \label{fig:xmcdhys}
\end{figure}

As a final measurement, hysteresis curves were obtained at the peak of the $L_3$ edge for both samples, as shown in Fig.~\ref{fig:xmcdhys}. In the case of \sampleMn, the hysteresis vanishes above \SI{90}{K}, in close agreement with our independent measurements on the bulk samples prior to sectioning ($T_\text{C}$ = \SI{96}{K}) using a magnetic property measurement system (MPMS). This lies \SI{10}{K} above the literature value of \SI{80}{K} \cite{Parkin1980} mentioned at the outset and indicates good sample quality, since off-stoichiometry drastically reduces the ferromagnetic transition temperature in 3\textit{d} ion intercalated TMDCs \cite{Hardy2015}. For \sampleFe, we measure a Curie temperature of \SI{\sim 110}{K}, which agrees reasonably well with our independent MPMS measurements on the bulk samples (in zero-field-cooled measurements, we witness a broad transition region extending from \SI{160}{K} to \SI{120}{K}). However, the measured transition temperature lies below the literature value of \SI{155}{K} \cite{Vannette2009} for phase pure \sampleFe, and is likely indicative of regions of off-stoichiometry \cite{Hardy2015}. This is also corroborated by our SAED results which indicate a degree of spatial inhomogeneity. Nonetheless, the expected high coercivity of the iron intercalated sample as compared to the manganese intercalated sample is clearly visible.

\section{Conclusion and Outlook}
In conclusion, we prepared free-standing thin films of magnetic ion intercalated transition metal dichalcogenides using ultramicrotoming techniques. Thicknesses ranging between \SI{30}{nm} and \SI{250}{nm} were achieved and characterized by transmission electron diffraction and XMCD spectroscopy. Using transmission electron diffraction, we witness the intercalant $2\times2$ crystallographic ordering in both materials, a characteristic of the intercalation concentration $y=1/4$. Using XMCD spectroscopy, we measured the orbital contribution to the total magnetization for the two materials.  Thin films facilitate true absorption measurements, and are preferred for quantitative measurements over fluorescence and electron yield geometries.  This has allowed us to verify the unquenched orbital moment in \sampleFe{}, as well as the fully quenched orbital moment in \sampleMn{}.

Looking ahead, we note that thin films of intercalated dichalcogenides have wide applicability in time-resolved X-ray and electron diffraction studies. In both cases, the sample thickness can be optimized to easily match the absorption depth of both pump and probe pulses, ensuring that measurements are performed on uniformly excited materials. With respect to time-resolved XMCD \cite{Radu, Bergeard}, we foresee the capability of measuring magnetization dynamics of `local moment' orbital and spin components on the intercalant site, as well as dynamics of itinerant conduction electrons of the host lattice (measured at a suitable host absorption edge). Due to the RKKY interaction, conduction electrons acquire a modest spin polarization which we expect exhibits different magnetization dynamics.  In the broader context of intercalated dichalcogenides, as one measures dynamics in a variety of members of this family, the orbital, spin, and conduction electron contribution to the total magnetization can be varied, providing a robust platform on which to study ultrafast magnetism.

Finally, we also envisage a range of time-resolved coherent X-ray scattering measurements to resolve the dynamics of magnetic domains on sub-picosecond timescales, and length scales of a few tens of nanometers. For the case of ultrafast transmission electron microscopy (UTEM) \cite{Lobastov2005,Feist2015}, thin magnetic films open the door to ultrafast Lorentz microscopy, thus providing a measure of magnetic domain motion and domain wall dynamics on timescales approaching femtoseconds, and spatial length scales of a few nanometers \cite{Park2010}.

\section*{Acknowledgments}
The authors thank P.~Pourhossein Aghbolagh and R.~Chiechi for discussions and instructions on the use of the ultramicrotome, as well as N.~Rubiano da Silva, M.~M\"oller, and J.~Momand for help with TEM diffraction images, and the HZB for the allocation of synchrotron radiation beamtime at BESSY II. Synchrotron work was supported by F.~Radu and T.~Kachel. TD, CR, and RT thank the German Academic Exchange Service (DAAD) for support within the framework of the U4 University Network during TD's stay in Groningen. XZ acknowledges support from the National Natural Science Foundation of China (Grant No.~11204312) for work conducted at the High Magnetic Field Lab (Hefei). The work at Rutgers University was supported by the NSF under Grant No.~NSF-DMREF-1233349, and the work at Postech was supported by the Max Planck POSTECH/KOREA Research Initiative Program (Grant No.~2011-0031558) through NRF of Korea funded by MSIP. The research leading to these results has received funding from the European Community's Seventh Framework Programme (FP7/2007-2013) under Grant Agreement No.~312284.


\bibliography{main}

\end{document}